\documentclass[aps,prb,showpacs,twocolumn,floats]{revtex4}
\usepackage{amssymb}

\usepackage{graphicx,psfig}
\usepackage{dcolumn}
\usepackage{bm}
\begin{document}

\bibliographystyle{prsty}

\title{
Magneto-elastic waves in crystals of magnetic molecules
\vspace{-1mm} }
\author{C. Calero, E. M. Chudnovsky, and D. A. Garanin}
 \affiliation{\mbox{Department of Physics and Astronomy,} \\
\mbox{Lehman College, City University of New York,} \\ \mbox{250
Bedford Park Boulevard West, Bronx, New York 10468-1589, U.S.A.}
\\ {\today}}
\begin{abstract}
We study magneto-elastic effects in crystals of magnetic
molecules. Coupled equations of motion for spins and sound are
derived and the possibility of strong resonant magneto-acoustic
coupling is demonstrated. Dispersion laws for interacting linear
sound and spin excitations are obtained for bulk and surface
acoustic waves. We show that ultrasound can generate inverse
population of spin levels. Alternatively, the decay of the inverse
population of spin levels can generate ultrasound. Possibility of
solitary waves of the magnetization accompanied by the elastic
twists is demonstrated.
\end{abstract}


\maketitle

\section{Introduction}
Crystals of molecular magnets are paramagnets \cite{book,dipolar}
that have the ability to maintain macroscopic magnetization for a
long time in the absence of the external magnetic field. This is a
consequence of the magnetic bi-stability of individual molecules
\cite{Sessoli} that in many respects behave as superparamagnetic
particles. The latter is due to a large spin (e.g., $S = 10$ for
Mn-12 and Fe-8) and high magnetic anisotropy of the molecules.
Together with quantization of spin energy levels this leads to a
distinctive feature of molecular magnets: A staircase hysteresis
curve in a macroscopic measurement of the magnetization.
\cite{Friedman}

Hybridization of electron paramagnetic resonance (EPR) with
longitudinal ultrasonic waves has been studied by Jacobsen and
Stevens \cite{JS} within a phenomenological model of
magneto-elastic interaction proportional to the magnetic field.
General theory of magneto-elastic effects on the phonon dispersion
and the sound velocity in conventional paramagnets has been
developed by Dohm and Fulde. \cite{DF} The advantage of molecular
magnets is that they, unlike conventional paramagnets, can be
prepared in a variety of magnetic states even in the absence of
the magnetic field. Spontaneous transitions between spin levels in
molecular magnets are normally due to the emission and absorption
of phonons. Interactions of molecular spins with phonons have been
studied in the context of magnetic relaxation,
\cite{Villain,GC-97,Loss,Comment} conservation of angular
momentum, \cite{EC-94,EC-Martinez,CGS} phonon Raman processes,
\cite{Raman} and phonon superradiance. \cite{SR} Parametric
excitation of acoustic modes in molecular magnets has been
studied. \cite{Tokman,Xie} It has been suggested that surface
acoustic waves can produce Rabi oscillations of magnetization in
crystals of molecular magnets. \cite{Rabi} In this paper we study
coupled dynamics of paramagnetic spins and elastic deformations at
a macroscopic level.

When considering magneto-elastic waves in paramagnets the natural
question is why the adjacent spins should rotate in unison rather
than behave independently. In ferromagnets the local alignment of
spins is due to the strong exchange interaction. Due to this
interaction the length of the local magnetization is a constant
throughout the ferromagnet. We shall argue now that a somewhat
similar quantum effect exists in a system of weakly interacting
two-level entities described by a fictitious spin $1/2$. Indeed,
since any product of Pauli matrices reduces to a single Pauli
matrix ${\sigma}_{\alpha}$, interaction of $N$ independent
two-state systems with an arbitrary field ${\bf A}({\bf r})$
should be linear on ${\sigma}_{\alpha}$,
\begin{equation}\label{gen-Ham}
{\cal{H}} = \sum_{n=1}^N g_{\alpha\beta}\sigma_{\alpha}^{(n)}
A_{\beta}({\bf r}_n) \;,
\end{equation}
where ${\bm \sigma}^{(n)}$ describes a two-state system located at
a point ${\bf r} = {\bf r}_n$. If ${\bf A}$ was independent of
coordinates, then the Hamiltonian (\ref{gen-Ham}) would reduce to
\begin{equation}
{\cal{H}} = g_{\alpha\beta}\Sigma_{\alpha}A_{\beta}\;,
\end{equation}
where
\begin{equation}
{\bf \Sigma} = \sum_{n=1}^N {\bm \sigma}^{(n)}
\end{equation}
is the total fictitious spin of the system. In this case the
interaction Hamiltonian would commute with ${\bf \Sigma}^2$, thus
preserving the length of the total fictitious ``magnetization''.
This observation is crucial for understanding Dicke superradiance:
\cite{Dicke} A system of independent two-state entities behaves
collectively in a field whose wavelength significantly exceeds the
size of the system. When the wavelength of the field is small
compared to the size of the system but large compared to the
distance between the two-state entities, the same argument can be
made about the rigidity of ${\bf \Sigma} = \sum {\bm
\sigma}^{(n)}$ summed up over the distances that are small
compared to the wavelength. Consequently, the system that has been
initially prepared in a state with all spins up, and then is
allowed to evolve through interaction with a long-wave Bose field,
should conserve the length of the local ``magnetization'' in the
same way as ferromagnets do.

The relevance of the above argument to the dynamics of magnetic
molecules interacting with elastic deformations becomes obvious
when only two spin levels are important. This is the case when the
low-energy dynamics of the molecular magnet is dominated by, e.g.,
tunnel split spin-levels or when the magneto-acoustic wave is
generated by a pulse of sound of resonant frequency. Recently,
experiments with surface acoustic waves in the GHz range have been
performed in crystals of molecular magnets. \cite{Alberto1} The
existing techniques, in principle, allow generation of acoustic
frequencies up to 100 GHz. \cite{Santos} This opens the
possibility of resonant interaction of generated ultrasound with
spin excitations. In this paper we study coupled magneto-elastic
waves in the ground state of a crystal of molecular magnets. We
derive equations describing macroscopic dynamics of sound and
magnetization and show that high-frequency ultrasound interacts
strongly with molecular spins when the frequency of the sound
equals the distance between spin levels. We obtain the dispersion
relation for magneto-elastic waves and show that non-linear
equations of motion also possess solutions describing solitary
waves of magnetization coupled to the elastic twists.

The paper is organized as follows. The model of spin-phonon
coupling is discussed in Section II where coupled magneto-elastic
equation are derived. Linear magneto-elastic waves are studied in
Section III where we obtain dispersion laws for bulk and surface
acoustic waves. Non-linear solitary waves are studied in Section
IV. Suggestions for experiments are made in Section V.

\section{Model of magneto-elastic coupling}
We consider a molecular magnet interacting with a local crystal
field described by a phenomenological anisotropy Hamiltonian
$\hat{\mathcal{H}}_A$.  The spin cluster is assumed to be more
rigid than its elastic environment, so that the long-wave crystal
deformations can only rotate it as a whole but cannot change its
inner structure responsible for the parameters of the Hamiltonian
$\hat{\cal{H}}_{A}$. This approximation should apply to many
molecular magnets as they typically have a compact magnetic core
inside a large unit cell of the crystal. In the presence of
deformations of the crystal lattice, given by the displacement
field ${\bf u}({\bf r})$, local anisotropy axes defined by the
crystal field are rotated by the angle
\begin{equation}\label{angle}
\delta {\bm \phi}({\bf r},t) = \frac{1}{2}\nabla \times {\bf
u}({\bf r},t)\,.
\end{equation}
As a consequence of the full rotational invariance of the system
(spins + crystal lattice), the rotation of the lattice is
equivalent to the rotation of the operator $\hat{\bf S}$ in the
opposite direction, which can be performed by the
$(2S+1)\times(2S+1)$ matrix in the spin space, \cite{CGS}
\begin{equation}\label{spin-rotation}
\hat{\bf S} \rightarrow \hat{R}^{-1}\hat{\bf S}\hat{R}, \qquad
\hat{R} = e^{i\hat{\bf S}\cdot \delta{\bm \phi}}\,.
\end{equation}
Therefore, the total Hamiltonian of a molecular magnet in the
magnetic field ${\bf B}$ must be written as
\begin{equation}\label{total Hamiltonian}
\hat{\mathcal{{H}}} = e^{-i\hat{\bf S}\cdot \delta{\bm
\phi}}\,\hat{\mathcal{{H}}}_A\,e^{i\hat{\bf S}\cdot \delta{\bm
\phi}} + \hat{\mathcal{H}}_Z + \hat{\mathcal{H}}_{ph}\,,
\end{equation}
where $\hat{\mathcal{H}}_A$ is the anisotropy Hamiltonian in the
absence of phonons, $\hat{\mathcal{H}}_Z = -g\mu_B {\bf B}\cdot
\hat{{\bf S}}$ is the Zeeman Hamiltonian and
$\hat{\mathcal{H}}_{ph}$ is the Hamiltonian of harmonic phonons.
The angle of rotation produced by the deformation of the lattice
is small, so one can expand Hamiltonian (\ref{total Hamiltonian})
to first order in the angle $\delta \phi$ and obtain
\begin{equation}
\hat{\mathcal{H}} \simeq \hat{\mathcal{H}}_0 +
\hat{\mathcal{H}}_{s-ph}\,,
\end{equation}
where $\hat{\mathcal{H}}_0$ is the Hamiltonian of non-interacting
spins and phonons
\begin{equation}
\hat{\mathcal{H}}_0 = \hat{\mathcal{H}}_S+
\hat{\mathcal{H}}_{ph}\,, \qquad \hat{\mathcal{H}}_S =
\hat{\mathcal{H}}_A + \hat{\mathcal{H}}_{Z}\,,
\end{equation}
and $\hat{\mathcal{H}}_{s-ph}$ is the spin-phonon interaction
term, given by
\begin{equation}\label{s-ph}
\hat{\mathcal{H}}_{s-ph}=  i\left[ \hat{\mathcal{H}}_A, \hat{{\bf
S}}\right]\cdot \delta{\bm \phi}\,.
\end{equation}

\subsection{Coupling of spins to the elastic twists}
For certainty, we consider a crystal of molecular magnets with the
anisotropy Hamiltonian
\begin{equation}\label{H-A}
\hat{\cal{H}}_{A} = -D\hat{S}_z^2 + \hat{V}\,,
\end{equation}
where $\hat{V}$ is a small term that does not commute with the
$\hat{S}_z$ operator. This term is responsible for the tunnel
splitting, $\Delta$, of the levels on resonance.

At low temperature and small magnetic field, $k_BT, g\mu_BB
\lesssim \Delta$, when the frequency of the displacement field
${\bf u}({\bf r})$ satisfies $\omega \ll 2DS/\hbar$, only the two
lowest states of $\hat{\mathcal{H}}_A$ are involved in the
evolution of the system. Thus, one can reduce the spin-Hamiltonian
of the molecular magnet to an effective two-state Hamiltonian in
terms of pseudospin-$1/2$ operators $\hat{\bf \sigma}_i$,
\begin{equation}\label{ham}
\hat{\mathcal{H}}_S^{(eff)} = -\frac{1}{2}\,(W{\bf e}_z + \Delta
{\bf e}_x)\cdot \hat{\bm \sigma}\, ,
\end{equation}
where $\hat{\sigma}_i$ are the Pauli matrices in the basis of the
$\hat{S}_z$-states close to the resonance between $|S\rangle$ and
$|-S\rangle$, and $W = E_{S} - E_{-S}$ is the energy difference
for the resonant states at $\Delta = 0$. The non-degenerate
eigenfunctions of $\hat{\mathcal{H}}_S^{(eff)}$ are
\begin{equation}\label{pm}
|\psi_{\mp}\rangle = \frac{1}{\sqrt{2}}\left( C_{\pm} |S\rangle
\mp C_{\mp}|-S\rangle \right)
\end{equation}
with
\begin{equation}
C_{\pm} = \sqrt{1\pm \frac{W}{\sqrt{\Delta^2 + W^2}}}\,.
\end{equation}
In terms of $|\psi_{\mp}\rangle$ the Hamiltonian (\ref{ham}) can
be written as
\begin{equation}
\hat{\mathcal{H}}_S^{(eff)} = -\frac{1}{2}\, \sqrt{W^2 +
\Delta^2}\, \hat{\tilde{\sigma}}_z\,,
\end{equation}
where $\hat{\tilde{\sigma}}_i$ are now the Pauli matrices in the
new basis $|\psi_{\pm}\rangle$, i.e., $\hat{\tilde{\sigma}}_z =
|\psi_+\rangle\langle \psi_+| - |\psi_-\rangle\langle \psi_-|$.
The projection of the spin-phonon interaction Hamiltonian
(\ref{s-ph}) onto this new two-state basis results in
\begin{equation}
\hat{\mathcal{H}}_{s-ph}^{(eff)} = \sum_{i,j = \pm} \langle
\psi_i|\hat{\mathcal{H}}_{s-ph}|\psi_j\rangle |\psi_i\rangle
\langle \psi_j| = \delta \phi_z S \Delta \hat{\tilde{\sigma}}_y\,,
\end{equation}
with $\hat{\tilde{\sigma}}_y = -i|\psi_+\rangle\langle \psi_-| +i
|\psi_-\rangle\langle \psi_+|$. The total Hamiltonian (\ref{total
Hamiltonian}) of a single molecular magnet becomes
\begin{eqnarray}\label{Heff}
\hat{\mathcal{H}}^{(eff)} & = & -\frac{1}{2} \,{\bf
b}^{(eff)}\cdot \hat{\tilde{\bm \sigma}} +
\hat{\mathcal{H}}_{ph}\,, \nonumber
\\ {\bf b}^{(eff)} & = & \sqrt{W^2 + \Delta^2}\,{\bf e}_z
-2\delta \phi_z S \Delta\,{\bf e}_y\,.
\end{eqnarray}
Here we have assumed that the perturbation introduced by the
spin-phonon interaction is much smaller than the perturbation
$\hat{V}$ producing the splitting $\Delta$, which will usually be
the case. Note also that $\Delta$ and $W$ can in general be made
${\bf r}$-dependent to account for possible inhomogeneities of the
crystal.

When considering magneto-elastic excitations we will need to know
whether they are accompanied by a non-zero local magnetization of
the crystal. For that reason it is important to have the magnetic
moment of the molecule,
\begin{equation}\label{mz}
m_z = g \mu_B \langle S_z \rangle\,,
\end{equation}
(with $g$ being the gyromagnetic ratio and $\mu_B$ being the Bohr
magneton), in terms of its wave function
\begin{equation}
|\Psi\rangle = K_+ |\psi_+\rangle + K_-|\psi_-\rangle\,,
\end{equation}
where $K_{\pm}$ are arbitrary complex numbers satisfying $|K_-|^2
+ |K_+|^2 = 1$. With the help of Eq.\ (\ref{pm}) one obtains
\begin{eqnarray}\label{Sz}
\frac{ \langle S_z\rangle}{S}  & = & \frac{W}{\sqrt{W^2 +
\Delta^2}} \left(|K_-|^2 - |K_+|^2\right) \nonumber \\
& + &\frac{\Delta}{\sqrt{W^2 + \Delta^2}}\left(K_+^*K_- +
K_+K_-^*\right)   \nonumber
\\ & = &
\frac{\Delta\langle \hat{\tilde{\sigma}}_x\rangle -W\langle
\hat{\tilde{\sigma}}_z \rangle}{\sqrt{W^2 + \Delta^2}} \,.
\end{eqnarray}

\subsection{Magneto-elastic equations}
We want to describe our system of $N$ spins in terms of the spin
field
\begin{equation}\label{4}
\hat{\bf n}({\bf r}) = \sum_{i}^N \hat{\tilde{\bm \sigma}}_i
\delta({\bf r} - {\bf r}_i)\,,
\end{equation}
satisfying commutation relations
\begin{equation}\label{5}
\left[\hat{n}_{\alpha}({\bf r}), \hat{n}_{\beta}({\bf r}') \right]
= 2i\epsilon_{\alpha \beta \gamma} \hat{n}_{\gamma}({\bf r})
\delta({\bf r} - {\bf r}')\,.
\end{equation}
In terms of this field the total Hamiltonian becomes
\begin{equation}\label{6}
\hat{\mathcal{H}} = -\frac{1}{2}\int d^3r\,\hat{\bf n}({\bf r})
\cdot {\bf b}^{(eff)}({\bf r})+\hat{\mathcal{H}}_{ph}\,.
\end{equation}
The classical pseudo-spin field ${\bf n}({\bf r}, t)$ can be
defined as
\begin{equation}
{\bf n}({\bf r},t) = \langle \hat{\bf n}({\bf r})\rangle\,,
\end{equation}
where $\langle ...\rangle$ contains the average over quantum spin
states and the statistical average over spins inside a small
volume around the point ${\bf r}$. If the size of that volume is
small compared to the wavelength of the phonon displacement field,
then, as has been discussed in the Introduction, ${\bf n}^2({\bf
r})$ should be approximately constant in time. According to
equations (\ref{mz}), (\ref{Sz}) and (\ref{4}), the magnetization
is given by
\begin{equation}\label{mag}
M_z({\bf r}) =g \mu_B S\; \frac{\Delta \, n_x({\bf r})-W\,
n_z({\bf r})}{\sqrt{W^2 + \Delta^2}} \,.
\end{equation}

The dynamical equation for the classical pseudo-spin field ${\bf
n}({\bf r},t)$ is
\begin{equation}
i\hbar\frac{\partial {\bf n}({\bf r},t)}{\partial t} =
\left\langle [\hat{\mathcal{H}},\hat{\bf n}] \right\rangle\,,
\end{equation}
which, with the help of Eq.\,(\ref{5}), can be written as
\begin{equation}\label{7}
\hbar\frac{\partial {\bf n}({\bf r},t)}{\partial t} =  {\bf
n}({\bf r},t) \times {\bf b}^{(eff)}({\bf r},t)\,.
\end{equation}
In this treatment we are making a common assumption that averaging
over spin and phonon states can be done independently. This
approximation is expected to be good in the long-wave limit.

The dynamical equation for the displacement field is
\begin{equation}\label{3}
\rho \frac{\partial^2 u_{\alpha}}{\partial t^2} =
\sum_{\beta}\frac{\partial \sigma_{\alpha \beta}}{\partial
x_{\beta}} \,,
\end{equation}
where $\sigma_{\alpha \beta} = {\partial h}/\partial e_{\alpha
\beta}$ is the stress tensor, $e_{\alpha \beta} = \partial
u_{\alpha}/\partial x_{\beta}$ is the strain tensor, $h$ is the
Hamiltonian density of the system in $ \hat{\mathcal{H}} = \int
d^3r\,h({\bf r})$, and $\rho$ is the mass density. Note that the
stress tensor has an antisymmetric part originating from the
magneto-elastic interaction in the Hamiltonian,
\begin{eqnarray}
\sigma_{\alpha \beta} &=&
\sigma_{\alpha \beta}^{(s)} + \sigma_{\alpha \beta}^{(a)}\,, \nonumber \\
\sigma_{\alpha \beta}^{(a)} &=& \frac{1}{2}S\Delta\, n_y({\bf r})
\epsilon_{z\alpha \beta}\,.
\end{eqnarray}
This implies that at each point ${\bf r}$ there is a torque per
unit volume,
\begin{equation}\label{9}
\tau_{\alpha}({\bf r}) = -\delta_{\alpha z} S\Delta\,n_y({\bf
r})\,,
\end{equation}
created by the interaction with the magnetic system. This effect
can be viewed as the local Einstein -- de Haas effect: Spin
rotation produces a torque in the crystal lattice due to the
necessity to conserve angular momentum. With the help of equations
(\ref{Heff}), (\ref{6}), and (\ref{3}), using standard results of
the theory of elasticity, one obtains
\begin{equation}\label{8}
\frac{\partial^2 {u}_{\alpha}}{\partial t^2} - c_t^2 {\bm
\nabla}^2 {u}_{\alpha} - (c_l^2-c_t^2){\nabla}_{\alpha}({\bm
\nabla}\cdot {\bf u}) = \frac{S\Delta}{2\rho} \,\epsilon_{z\alpha
\beta} {\nabla}_{\beta}n_y\,,
\end{equation}
where $c_l$ and $c_t$ are velocities of longitudinal and
transverse sound. The source of deformation in the right hand side
of this equation is due to the above-mentioned torque generated by
the spin rotation.

Equations (\ref{7}) and (\ref{8}) describe coupled motion of the
pseudospin field ${\bf n}({\bf r},t)$ and the displacement field
${\bf u}({\bf r},t)$. It is easy to see from these equations that
in accordance with the argument presented in the Introduction
$n_x^2 + n_y^2 + n_z^2$ is independent of time. It may,
nevertheless, depend on coordinates, reflecting the structure of
the initial state. In this paper we study cases in which the
crystal of molecular magnets was initially prepared in the ground
state ${\bf n} = n_0{\bf e}_z$ with $n_0$ being the concentration
of magnetic molecules. In this case the dynamics of ${\bf n}({\bf
r})$ described by equations (\ref{7}) and (\ref{8}) reduces to its
rotation, with the length of ${\bf n}({\bf r})$ being a constant
$n_0$. Remarkably, this situation is similar to a ferromagnet,
despite the absence of the exchange interaction.

\section{Linear magneto-elastic waves}

\subsection{Bulk waves}
For magnetic molecules whose magnetic cores are more rigid than
their environments, only the transverse part of the displacement
field (with $\nabla \cdot {\bf u}({\bf r}) = 0$) interacts with
the magnetic degrees of freedom. This is a consequence of the fact
that the elastic deformation produced by the rotation of local
magnetization is a local twist of the crystal lattice, required by
the conservation of angular momentum. Let us consider then a
transverse plane wave propagating along the X-axis. From
Eqs.\,(\ref{7}) and (\ref{8}) one obtains
\begin{eqnarray}\label{full-sys}
& & \frac{\partial^2 u_{y}}{\partial t^2} -c_t^2\frac{\partial^2
u_{y}}{\partial x^2} = -\frac{S\Delta}{2\rho}\frac{\partial
n_y}{\partial x} \label{elasticeq} \nonumber \\
& & \hbar \frac{\partial n_x}{\partial t} = n_y \sqrt{W^2 +
\Delta^2} - n_z S\Delta  \frac{\partial u_y}{\partial x} \nonumber  \\
& & \hbar \frac{\partial n_y}{\partial t} = -n_x \sqrt{W^2 +
\Delta^2} \label{neq} \nonumber \\
& & \hbar \frac{\partial n_z}{\partial t} = S\Delta n_x
\frac{\partial u_y}{\partial x}\;.
\end{eqnarray}

We shall study linear waves around the ground state
$|\psi_+\rangle$ corresponding to  $n_z = n_0, n_{x,y} =0, u_y =
0$. The perturbation around this state results in nonzero
$n_{x,y}$ and $u_y$. Linearized equations of motion are
\begin{eqnarray}\label{sys-n}
& & \frac{\partial^2 u_{y}}{\partial t^2} -c_t^2\frac{\partial^2
u_{y}}{\partial x^2} = -\frac{S\Delta}{2\rho}\frac{\partial
n_y}{\partial x} \nonumber \\
& & \hbar \frac{\partial n_x}{\partial t} = n_y \sqrt{W^2 +
\Delta^2} - S\Delta n_0 \frac{\partial u_y}{\partial x} \nonumber \\
& & \hbar \frac{\partial n_y}{\partial t} = -n_x \sqrt{W^2 +
\Delta^2} \;.
\end{eqnarray}
For $u_y, n_{x,y} \propto \exp(iqx - i\omega t)$, the above
equations become
\begin{eqnarray}\label{system}
& & (\omega^2 - c_t^2q^2)u_y - iq\frac{S\Delta}{2\rho}n_y = 0 \nonumber \\
& & iq\frac{n_0 S\Delta  \sqrt{W^2 +\Delta^2}}{\hbar^2}u_y +
\left(\omega^2 -\frac{W^2 + \Delta^2}{\hbar^2}\right)n_y = 0\,.
\nonumber
\\
\end{eqnarray}
\begin{figure}
\unitlength1cm
\begin{picture}(20,5)
\centerline{\psfig{file=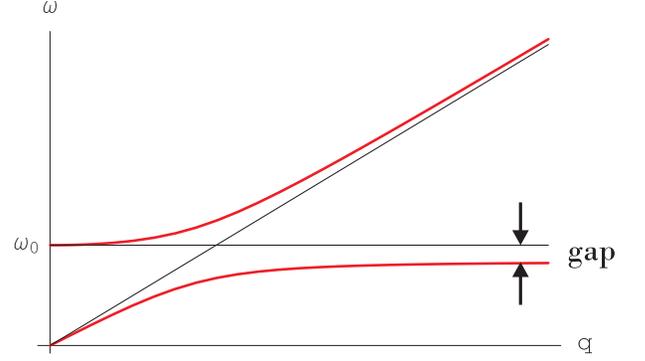,width=8cm}}
\end{picture}
\caption{\label{spectrum-Pi} Interacting sound and spin modes.
Notice the gap below spin resonance $\omega_0$.}
\end{figure}
The spectrum of coupled excitations is given by
\begin{equation}\label{dis}
(\omega^2 - c_t^2q^2)\left(\omega^2 - \frac{W^2 +
\Delta^2}{\hbar^2}\right) = \frac{n_0 S^2\Delta^2\sqrt{W^2
+\Delta^2}}{2 \rho \hbar^2} q^2\;.
\end{equation}
In the vicinity of the resonance,
\begin{equation}\label{res}
c_tq_0 = \frac{\sqrt{W^2 + \Delta^2}}{\hbar} \equiv \omega_0\;,
\end{equation}
one can write
\begin{equation}\label{split}
\omega = \omega_0(1 + \delta)\;
\end{equation}
with $\delta$ to be determined by the dispersion relation.
Substituting equations (\ref{res}) and (\ref{split}) into Eq.\
(\ref{dis}), one obtains
\begin{equation}\label{me-split}
\delta = \pm \sqrt{\frac{n_0 S^2 \Delta^2}{8\rho c_t^2 \hbar
\omega_0}}\;,
\end{equation}
that describes the splitting of two coupled modes at the
resonance. The repulsion of elastic and spin modes is illustrated
in Fig. \ref{spectrum-Pi}. The relative splitting of the modes
reaches maximum at $W = 0$ ($\hbar \omega_0 = \Delta$):
\begin{equation}\label{delta}
2|\delta_{max}| = \sqrt{\frac{n_0 S^2 \Delta}{2\rho c_t^2 }} = S
\sqrt{\frac{\Delta}{2{\rm M} c_t^2 }} \;,
\end{equation}
where ${\rm M} = \rho/n_0$ is the mass of the volume containing
one molecule of spin $S$. Notice also another consequence of Eq.\
(\ref{dis}): The presence of the energy gap below $\omega_0 =
\sqrt{W^2 + \Delta^2}/\hbar$ (see Fig. \ref{spectrum-Pi}). The
value of the gap follows from Eq.\ (\ref{dis}) at large $q$. It
equals $2\delta^2\omega_0$. This effect is qualitatively similar
to the one obtained in Ref. \onlinecite{JS} from an {\it ad hoc}
model of spin-phonon interaction. In contrast with that model our
results for the splitting of the modes and for the gap do not
contain any unknown interaction constants as they are uniquely
determined by the conservation of the total angular momentum (spin
+ crystal lattice).

According to equations (\ref{system}) and (\ref{dis}) the Fourier
transforms of $n_y$ and $u_y$ are related through
\begin{equation}
\frac{n_y}{n_0} = iS \frac{\omega_0^2}{\omega_0^2 -
\omega^2}\;\frac{\Delta}{\hbar \omega_0} \; q u_y\,.
\end{equation}
Due to the condition of the elastic theory $q u_y \ll 1$, the
absolute value of the ratio $n_y/n_0$ is generally small, unless
$\omega$ is close to $\omega_0$. This means that away from the
resonance the sound cannot significantly change the population of
excited spin states. At the magneto-elastic resonance,
substituting equations (\ref{split}) and (\ref{me-split}) into the
above equation, one obtains:
\begin{equation}\label{res-mix}
\frac{|n_y|_{res}}{n_0} = \left(\frac{2{\rm
M}\omega_0}{\hbar}\right)^{1/2}|u_y|\,.
\end{equation}
Although this relation is valid only at $|n_y| \ll n_0$, it allows
one to estimate the amplitude of ultrasound that will
significantly affect populations of spin states. We shall postpone
the discussion of this effect until Section \ref{Discussion}.
Meantime let us compute the magnetization generated by the linear
elastic wave, $u_y = u_0 \cos [q_0(x - c_t t)])$, in resonance
with our two-state spin system. The last of Eqs.\ (\ref{sys-n})
yields $n_x = i(\omega/\omega_0)n_y$. Then, with the help of Eq.\
(\ref{mag}) and Eq.\ (\ref{res-mix}) one obtains
\begin{equation}\label{mag-sound}
M_z = g \mu_B S \frac{\Delta}{\hbar \omega_0}\left(\frac{2{\rm
M}c_t^2}{\hbar \omega_0}\right)^{1/2}q_0 u_0 \cos [q_0(x - c_t
t)]\,.
\end{equation}

So far we have investigated coupled magneto-elastic waves in the
vicinity of the ground state, $n_z = n_0$. Eqs.\ (\ref{full-sys})
also allow one to obtain the increment, $\Gamma$, of the decay of
the unstable macroscopic state of the crystal, $n_z =-n_0$, in
which all molecules are initially in the excited state
$|\psi_-\rangle$. In fact, the result can be immediately obtained
from equations (\ref{sys-n}) -- (\ref{dis}) by replacing $n_0$
with $-n_0$. It is then easy to see from Eq.\ (\ref{dis}) that in
the vicinity of the resonance the frequency acquires an imaginary
part that attains maximum at the resonance where
\begin{equation}\label{decay}
\omega = \omega_0(1 \pm i|\delta|)\,.
\end{equation}
The mode growing at the rate $\Gamma = \omega_0|\delta|$
represents the decay of $|\psi_-\rangle$ spin states into
$|\psi_+\rangle$ spin states, separated by energy $\hbar
\omega_0$. This decay is accompanied by the exponential growth of
the amplitude of ultrasound of frequency $\omega_0$.

\subsection{Surface waves}
\begin{figure}
\unitlength1cm
\begin{picture}(15,3)
\centerline{\psfig{file=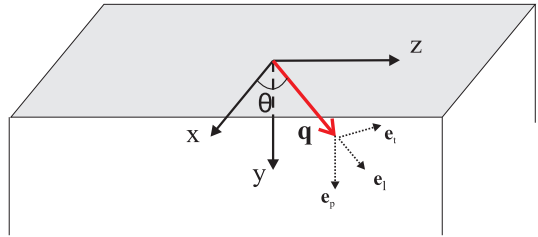,width=7cm}}
\end{picture}
\caption{\label{geometry} Geometry of the problem with surface
acoustic waves. }
\end{figure}

Magneto-elastic coupling in crystals of molecular magnets can be
studied with the help of surface acoustic waves (see Discussion).
To describe the surface waves we chose a geometry in which the
surface of interest is the $XZ$-plane and the solid extends to
$y>0$ with waves running along the direction that makes an angle
$\theta$ with the $X$-axis, see Fig.\,\ref{geometry}.
As usual \cite{LL} we assume that the displacement field ${\bf
u}({\bf r},t)$ and the components $n_x({\bf r}, t), n_y({\bf r},
t)$ have the form
\begin{equation}
A = A_0 e^{-\alpha y}e^{i(q_xx + q_zz)}e^{-i\omega t}\,.
\end{equation}
It is convenient to express the components of the displacement
field in the coordinate system defined by $({\bf e}_l, {\bf e}_t,
{\bf e}_p)$, see Fig.\,\ref{geometry},
\begin{eqnarray}
u_x & = & u_l\cos\theta - u_t \sin\theta \nonumber \\
u_y & = & u_p\nonumber \\
u_z & = & u_l \sin\theta + u_t\cos\theta\,.
\end{eqnarray}
Equations of motion for $u_l$, $u_t$, and $u_p$ follow from Eq.\
(\ref{8}):
\begin{eqnarray}\label{11}
\left[\omega^2 + c_t^2(\alpha^2 - q^2) \right]u_t &+&
\frac{S}{2\rho}\,\alpha \Delta \sin\theta n_y = 0 \nonumber \\
\left[\omega^2 + c_t^2\alpha^2 - c_l^2q^2\right]u_l
 &-& i \alpha q(c_l^2 - c_t^2)u_p
\nonumber \\ &-& \frac{S}{2\rho}\,\alpha \Delta \cos\theta n_y=0 \nonumber \\
\left[\omega^2 + c_l^2\alpha^2 - c_t^2q^2\right]u_p &-& i \alpha
q(c_l^2 - c_t^2)u_l \nonumber \\ &-& \frac{i S
}{2\rho}\, \Delta q \cos\theta \,n_y =0\,. \nonumber \\
\end{eqnarray}
It is easy to see that for $\theta \neq k\pi\,,\;\; k = 0, 1, 2...
$ and $n_y \neq 0$, the transverse component $u_t$ cannot be zero,
contrary to the case of Rayleigh waves. This is the signature of
magneto-elastic coupling.

As in the analysis of bulk waves, we shall study the linear waves
around the ground state corresponding to the pseudospin field
polarized in the $Z$-direction, $n_z = n_0, n_{x,y} =0$. The
excitations above this state are described by Eqs.\ (\ref{7}),
which become
\begin{eqnarray}
-i\hbar\omega n_{x} & = & S\Delta\, \left[-\alpha (u_l \cos\theta
- u_t\sin\theta) -i q_{\parallel} \cos\theta u_p \right] \nonumber
\\ &+& \sqrt{W^2 +
\Delta^2} \,n_{y}\nonumber \\
-i\hbar\omega n_{y} & = & -\sqrt{W^2 + \Delta^2}\, n_{x}\;.
\end{eqnarray}
Substitution of these two equations into Eqs.\,(\ref{11}) leads to
a homogeneous system of algebraic equations for $u_l$, $u_t$, and
$u_p$, that have a non-zero solution only if its determinant
equals zero. From this condition we obtain three values of the
coefficient $\alpha$ that describe the decay of the wave away from
the surface:
\begin{eqnarray}
\alpha_1 &=& \sqrt{q^2 - \frac{\omega^2}{c_l^2}}\,, \quad \alpha_2
=\sqrt{q^2 - \frac{\omega^2}{c_t^2}}\,,\nonumber \\
\alpha_3 &=& \sqrt{\frac{c_t^2q^2 - \omega^2 + \eta q^2
\cos^2\theta}{\eta + c_t^2}}\,,
\end{eqnarray}
where
\begin{equation}
\eta \equiv \frac{ S^2\Delta^2 \sqrt{W^2 + \Delta^2}}{2{\rm
M}\left[\hbar^2\omega^2 - (W^2 + \Delta^2)\right]}\,.
\end{equation}
Note that if there are no spins ($S = 0$), then $\alpha_3 =
\alpha_2$ and one obtains decay coefficients for ordinary Rayleigh
waves.

The general plane wave solution for the components of the
displacement field can be written as
\begin{equation}
u_i = \sum_{k =1}^3u_{i0}^{(k)} e^{-\alpha_k y}e^{i(q_xx +
q_zz)}e^{-i\omega t}\,,
\end{equation}
where $u_{i0}^{(k)}$ is the amplitude corresponding to each
$\alpha_k$ and $i=l, t,p$. For each $k$, the amplitudes
$u_{l0}^{(k)}, u_{t0}^{(k)}, u_{p0}^{(k)}$ are related through
Eqs.\,(\ref{11}) (there are two independent equations, so we can
express, e.g., $u_{t0}^{(k)}, u_{p0}^{(k)}$ in terms of
$u_{l0}^{(k)}$). Therefore, there still are three unknowns, say
$u_{l0}^{(1)}, u_{l0}^{(2)}, u_{l0}^{(3)}$. The boundary
conditions for the stress tensor at the surface, $\sigma_{iy}|_{{y
= 0}} = 0$, provide a system of homogeneous equations for
$u_{l0}^{(1)}, u_{l0}^{(2)}$ and $u_{l0}^{(3)}$, whose determinant
must be zero to allow for non-trivial solution. From this last
condition we obtain the dispersion relation for surface
magneto-elastic waves:
\begin{eqnarray}\label{dispersion}
&&- 4q^2\sqrt{q^2 - \frac{\omega^2}{c_l^2}}\,\Bigg[\left(q^2 -
\frac{\omega^2}{c_t^2}\right)^{3/2}\sin^2\theta  -
\frac{\omega^2}{c_t^2}\cos^2\theta \times \nonumber \\ & &
\sqrt{\frac{q^2S^2\Delta^2\omega_0\cos^2\theta + 2{\rm M}
c_t^2\hbar(\omega^2 - \omega_0^2)(q^2 -
\omega^2/c_t^2)}{S^2\Delta^2\omega_0 + 2{\rm M} c_t^2\hbar(
\omega^2 - \omega_0^2)}} \,\Bigg] \nonumber
\\ &&+ \left(2q^2 - \frac{\omega^2}{c_t^2}\right)^2\left(
q^2\sin^2\theta - \frac{\omega^2}{c_t^2}\right) =  0\,.
\end{eqnarray}
This equation should be solved numerically to obtain the
dispersion law for magneto-elastic modes. Qualitatively, the
repulsion of the modes is similar to the one shown in Fig.
\ref{spectrum-Pi}.

\section{Non-linear magneto-elastic waves}
An interesting feature of Eqs.\ (\ref{full-sys}) is the existence
of transverse non-linear plane wave solutions of the form $u_i =
u_i(x - vt), n_i = n_i(x-vt)$. For such a choice, Eq.\
(\ref{elasticeq}) gives
\begin{equation}
\frac{du_y}{d\bar{x}} = \frac{S\Delta}{2 \rho (c_t^2 - v^2)}
n_y\;,
\end{equation}
where $\bar{x} \equiv x - vt$ and the constant of integration was
put zero assuming that there is no $du_y/d\bar{x}$ independent
from $n_y$. Substituting this into the equations of motion for
${\bf n}$, Eqs.\,(\ref{neq}), one obtains
\begin{eqnarray}
& & -\frac{dn_x}{d\xi} = n_y - \gamma n_y n_z \nonumber \\
& & -\frac{dn_y}{d\xi} = - n_x \label{nneq}\\
& & -\frac{dn_z}{d\xi} = \gamma n_xn_y\;,\nonumber
\end{eqnarray}
where
\begin{equation}\label{gamma}
\xi \equiv \frac{\bar{x}\sqrt{W^2 + \Delta^2}}{\hbar v}\, \qquad
\gamma \equiv \frac{S^2\Delta^2}{2 \rho (c_t^2 - v^2)\sqrt{W^2 +
\Delta^2}}
\end{equation}
The system of Eqs.\,(\ref{nneq}) can be reduced to
\begin{eqnarray}
n_z & = & C - \frac{1}{2}\gamma n_y^2 \\
\frac{d^2n_y}{d\xi^2} & = & -n_y\left(1-\gamma C +
\frac{1}{2}\gamma^2 n_y^2\right)\;,\label{nleq}
\end{eqnarray}
where $C$ is a constant of integration. The first integral of the
last differential equation is
\begin{equation}\label{first integral}
\frac{1}{2}\left( \frac{dn_y}{d\xi} \right)^2 =
-\frac{1}{2}(1-\gamma C)n_y^2 - \frac{\gamma^2}{8}n_y^4 +A \ge 0
\,,
\end{equation}
where $A$ is another integration constant.

We are interested in real bounded solutions of Eq.\,(\ref{nleq})
with $n_y$ vanishing at $ x- vt \rightarrow \pm \infty$, so that
the integration constant $A$ must be zero. In this case, for the
right hand side of Eq.\,(\ref{first integral}) to be positive we
must have $1-\gamma C <0$. Then, the solution of Eq.\,(\ref{nleq})
is
\begin{equation}
n_y(\xi) = \frac{\sqrt{\gamma C -1}\, e^{\pm\sqrt{\gamma C
-1}(\xi-\xi_0)}}{\gamma + \gamma e^{\pm2\sqrt{\gamma C
-1}(\xi-\xi_0)}}\,.
\end{equation}
From the equations
\begin{equation}\label{nx}
n_x = \frac{dn_y}{d\xi} \,,\qquad n_z = C - \frac{1}{2}\gamma
n_y^2
\end{equation}
one determines with the help of the condition $n_x^2 + n_y^2 +
n_z^2 =n_0^2$ that $C =  \pm n_0$ . Therefore, $\gamma$ must
satisfy $|\gamma|
>1/n_0$ for the equation (\ref{nleq}) to have a solution satisfying
the conditions specified above. Setting the reference point $\xi_0
= 0$ one obtains
\begin{equation}\label{ny}
n_y(\xi) =  \pm\frac{2}{|\gamma| }\sqrt{|\gamma|n_0-1}\;
{\text{sech}} \left[\sqrt{|\gamma |n_0-1}\; \xi \right]\,,
\end{equation}
so that
\begin{equation}\label{soliton}
n_z(\xi) =  \pm 1 \mp 2\,\frac{|\gamma |n_0 -1}{|\gamma| }\;
{\text{sech}} ^2\left[\sqrt{|\gamma |n_0-1}\;\xi \right]\,.
\end{equation}
In these formulas, the upper sign corresponds to $\gamma >0$ and
the lower sign to $\gamma <0$.

Eq.\ (\ref{soliton}) describes a solitary wave of a characteristic
width
\begin{equation}\label{width}
l_0 \sim \frac{1}{\sqrt{|\gamma|n_0-1}}\frac{\hbar v}{\sqrt{W^2 +
\Delta^2}}\,,
\end{equation}
travelling at a speed $v$. The parameter $\gamma$ given by Eq.\
(\ref{gamma}) is determined by $v$, which is the only free
parameter of the soliton. The magnetization inside the soliton is
given by Eq.\ (\ref{mag}) with $n_x$ and $n_z$ defined by
equations (\ref{nx}) -- (\ref{soliton}). At, e.g., $W =0$
\begin{eqnarray}
&& M_z = \mp g\mu_B \frac{2S(|\gamma|n_0 -1)}{|\gamma|} \times
\nonumber
\\ && \times \text{sech}\left[ \sqrt{|\gamma|n_0 -1}\,\xi
\right] \tanh\left[\sqrt{|\gamma|n_0 -1}\,\xi\right]\,.\nonumber \\
\end{eqnarray}
\begin{figure}
\unitlength1cm
\begin{picture}(20,5)
\centerline{\psfig{file=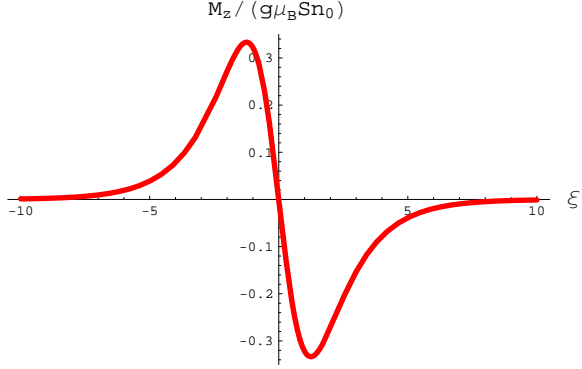,width=8cm}}
\end{picture}
\caption{\label{magnetization} Magnetization inside the soliton as
a function of $\xi$ for $W = 0$. }
\end{figure}
The condition
\begin{equation}
n_0|\gamma| = \frac{1}{|1 - v^2/c_t^2|}\,\frac{S^2\Delta}{2 {\rm
M}c_t^2 }\,\frac{\Delta}{\sqrt{W^2 + \Delta^2}} > 1\,
\end{equation}
requires $v$ to be very close to the speed of sound $c_t$. This is
a consequence of $\Delta$ being very small compared to ${\rm
M}c_t^2$.  Note that the maximal value of the magnetization inside
the soliton,
\begin{equation}
|M_z| = g\mu_B S\left(n_0 - \frac{1}{|\gamma|}\right)\,,
\end{equation}
is, in general, of the order of saturation magnetization $M_0 =
g\mu_B S n_0$. We should also note that although the above
non-linear solution of the equations of motion formally allows $v$
to be both slightly lower or slightly higher than $c_t$, the
supersonic soliton should be unstable with respect to Cherenkov
radiation of sound waves.

\section{Discussion}\label{Discussion}

Eq.\ (\ref{delta}) provides the splitting of the bulk sound
frequency in a magnetized crystal of magnetic molecules in the
vicinity of the resonance between sound and spin levels. At a zero
field bias ($W = 0$) the resonant condition, $\Delta = \hbar c_t
q$, should be easily accessible at low $\Delta$. However, the
splitting given by Eq.\ (\ref{delta}) will be very small unless
$\Delta$ is in the GHz range or higher. Such a large $\Delta$ will
be also beneficial for decreasing inhomogeneous broadening of
$\Delta$ and for insuring low decoherence of quantum spin states.
Surface acoustic waves can, in principle, be generated up to
$100$GHz \cite{Santos}. They may also be easier to use for the
observation of the discussed splitting. By order of magnitude it
will still be given by Eq.\ (\ref{delta}). Substituting into this
equation $S = 10$, $\Delta \sim 0.1\,$K (frequency $f$ in the GHz
range), ${\rm M}c_t^2 \sim 10^5$K, one obtains ${\delta}_{max}
\sim 10^{-2}$. This will be observable if the quality factor of
ultrasound in the GHz range exceeds $100$. The magneto-elastic
nature of the splitting can be confirmed through its dependence on
the angle between the wave vector and the easy magnetization axis
of the crystal, see Sec. III-B. Observation of the gap,
$2\delta^2\omega_0$, in the excitation spectrum (see Fig.
\ref{spectrum-Pi}) will be more challenging. For practical values
of $\delta$ the gap is likely to be small compared to the width of
the spin resonance and the width of the ultrasonic mode in the GHz
range.

Eq.\ (\ref{res-mix}) shows that at ${\rm M} \sim 10^{-21}$g and
$\omega_0 \sim 10^{10}$s$^{-1}$ ultrasound of amplitude $u_0 \sim
0.1\,$nm will significantly affect population of spin levels.
Moreover, it will result in the oscillating magnetization of large
amplitude, Eq.\ (\ref{mag-sound}). We have also demonstrated that
one can prepare the crystal in the excited spin state and generate
ultrasound due to the decay of the population of that state. This
result is another confirmation of the phonon laser effect
suggested in Ref. \onlinecite{SR}. Equations (\ref{decay}) and
(\ref{delta}) show that at $\omega_0 \sim 10^{10}$s$^{-1}$ the
amplitude of the sound wave may grow at the rate as high as
$\Gamma \sim 10^{8}$s$^{-1}$. Magneto-elastic effects studied in
this paper should be sensitive to the decoherence of spin states.
However, when the oscillation of spin population is driven by the
external acoustic wave, the latter should force the phase
coherence upon the spin system. To provide the resonance
condition, the broadening of the level splitting due to disorder
and dipolar fields should be small compared to $ \Delta$. If it is
not, the tunnel splitting, $\Delta$, should be increased by
applying a sufficiently large transverse magnetic field.

One fascinating prediction of our theory is the existence in
molecular magnets of solitary waves of the magnetization reversal
coupled to elastic twists. Such waves have quantum origin as they
are related to the quantum splitting of spin-up and spin-down
states. They can be ignited in experiment that starts with all
molecules in the ground state. Such a state of the crystal has
zero magnetization as the molecules are in a superposition of
spin-up and spin-down states. The soliton discussed above is
characterized by a narrow region of a large non-zero magnetization
that propagates through the solid with the velocity close to the
speed of transverse sound. It can be generated by, e.g., a
localized pulse of the magnetic field or by a localized mechanical
twist, and detected through local measurements of the
magnetization. In general the width of the soliton, given by Eq.\
(\ref{width}), is of order of the wavelength of sound of frequency
$\sqrt{W^2 + \Delta^2}/\hbar$, though wider solitons are allowed
if $|\gamma| n_0 \rightarrow 1$. In experiment this width should
depend on the width of the field pulse or the size of the twisted
region that generates the soliton.

\section{Acknowledgements}

This work has been supported by the NSF Grant No. EIA-0310517.

\end{document}